\documentclass[aps,prc,reprint,a4paper,superscriptaddress]{revtex4-1}
\usepackage{amsmath, amssymb, geometry, graphicx}
\usepackage{color,soul,lineno}
\begin{document}
\title{Measurements of the Neutron-Proton and Neutron-Carbon Total Cross Section from 150 to 800 keV}
\date{\today}
\author{B.~H.~Daub}
\email[Electronic address: ]{daubb@berkeley.edu}
\altaffiliation[Currently at ]{University of California, Berkeley, Berkeley, CA}
\affiliation{Laboratory for Nuclear Science and Department of Physics, Massachusetts Institute of Technology, Cambridge, MA 02139}
\author{V.~Henzl}
\altaffiliation[Currently at ]{Los Alamos National Laboratory, Los Alamos, NM}
\affiliation{Laboratory for Nuclear Science and Department of Physics, Massachusetts Institute of Technology, Cambridge, MA 02139}
\author{M.~A.~Kovash}
\affiliation{Department of Physics and Astronomy, University of Kentucky, Lexington, KY 40506}
\author{J.~L.~Matthews}
\affiliation{Laboratory for Nuclear Science and Department of Physics, Massachusetts Institute of Technology, Cambridge, MA 02139}
\author{Z.~W.~Miller}
\author{K.~Shoniyozov}
\author{H.~Yang}
\affiliation{Department of Physics and Astronomy, University of Kentucky, Lexington, KY 40506}

\begin{abstract}
There have been very few measurements of the total cross section for  $np$ scattering below 500 keV.  In order to differentiate among $NN$ potential models, improved cross section data between 20 and 600 keV are required.  We measured the $np$ and $n$C total cross sections in this energy region by transmission; a collimated neutron beam was passed through CH$_2$ and C samples and transmitted neutrons were detected by a BC-501A liquid scintillator.  Cross sections were obtained with a precision of 1.1-2.0\% between 150 and 800~keV using ratios of normalized neutron yields measured with and without the scattering samples in the beam. In energy regions where they overlap, the present results are consistent with existing precision measurements, and fill in a significant gap in the data between $E_n = 150$ and 500~keV.
\end{abstract}

\maketitle
\section{Introduction}

Nucleon-nucleon interactions and $NN$ potential models are an important representation of the strong interaction and a component of the theory of nuclei.  Measurements of the two-body interaction in $pp$ and $np$ reactions have been applied to studies of the structure and dynamics of light nuclei as well as testing the charge, spin, and isospin dependence of the strong nuclear force~\cite{RevModPhys.70.743}.

Additionally, $np$ scattering data are applicable to the detection of neutrons in organic scintillators below 2 MeV, where the primary mechanism for depositing energy in the scintillator is $np$ elastic scattering.  Therefore, determining the efficiency of a neutron detector using either Monte Carlo or analytical methods requires accurate knowledge of the $np$ elastic cross section.  Since the $np\rightarrow d\gamma$ cross section is typically 5 orders of magnitude smaller than the elastic cross section in this energy range~\cite{hale04}, measurement of the total $np$ cross section effectively yields the elastic cross section.

\begin{figure*}
\centering
\includegraphics[width=\textwidth]{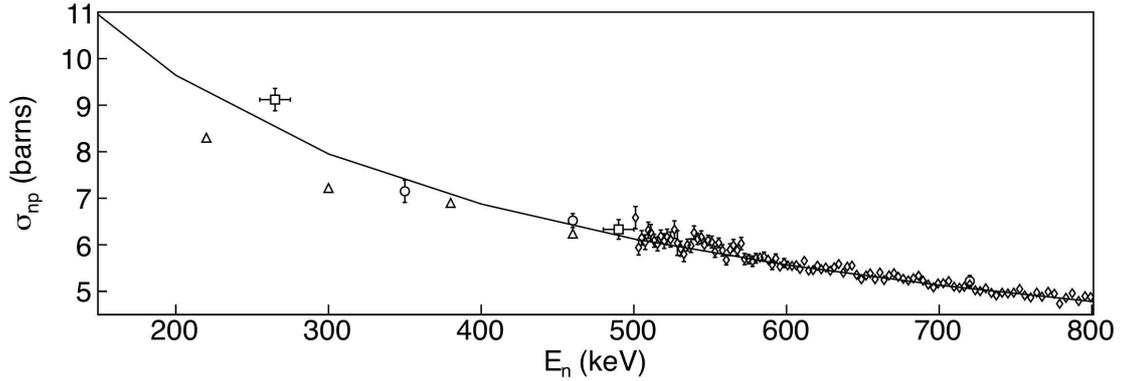}
\caption{Existing $np$ total cross section measurements for neutron energies from 150 to 800 keV, measured by (squares) Frisch~\cite{frisch46}, (circles) Bailey et. al.~\cite{bailey46}, (triangles) Bretscher et. al.~\cite{bretscher50}, and (diamonds) Clement et. al.~\cite{Clement197251} plotted with the ENDf/B-VII.1 tabulation (solid line) from Hale et. al.~\cite{hale04}. }
\label{fig:1}
\end{figure*}

However, the relevant data~\cite{frisch46,bailey46,bretscher50,Clement197251}, plotted with the ENDf/B-VII.1 tabulation~\cite{hale04} in figure~\ref{fig:1}, reveal a lack of precise measurements between 150 and 500~keV.
Previous measurements all employed the method of neutron transmission to determine the total cross section, using polyethylene and carbon samples to account for the carbon contribution and obtain the $np$ cross section.  We focused on the range between 150 and 800 keV, covering the region below 500 keV where there are very limited data.

\section{Experimental Procedure}

We measured the total cross section for neutron scattering from polyethylene and pure carbon samples using the method of neutron transmission.  The neutron beam were produced via the $^7$Li$(p,n)^7$Be reaction, with a 1.875 MHz pulsed proton beam produced by the Van de Graaff accelerator located at the University of Kentucky.  The Q-value for this reaction is -1.644 MeV~\cite{liskien75}.  The neutron production target was composed of 20 k$\mathring{\text{A}}$ thick LiF on a tantalum backing; this thickness was chosen such that 2.25 MeV protons would lose approximately 50 keV before exiting the target.  Using a thin target limited the spread of neutron energies and reduced the backgrounds.

\begin{figure}
\includegraphics[width=7cm]{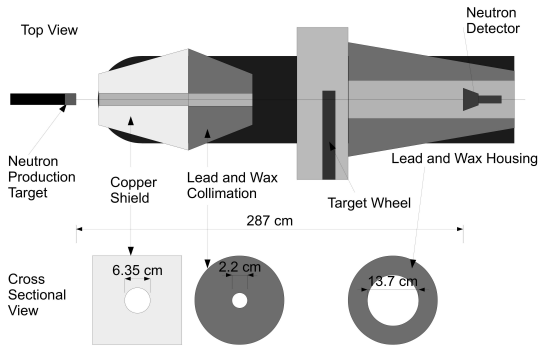}
\caption{Detector and shielding configuration.}
\label{fig:2}
\end{figure}

The neutrons were collimated by a copper shield, with a 6.35 cm opening and 50.8 cm thickness.  The beam was additionally defined by a wax collimator with an opening matched to that of the copper shield and tapered down to a 2.2 cm opening.  The sample was placed in the beamline past this collimation.  The neutrons were detected by a 13.7 cm diameter BC-501A liquid scintillator, which was housed in a lead and wax shield, with a 4.4 cm lead lining in the internal aperture and a 7 cm wax lining around the lead.  The experimental setup is shown in figure~\ref{fig:2}.
The geometry of the collimation resulted in a beam size smaller than both the samples and the neutron detector.  This was tested by performing several runs where a sample was shifted by several centimeters off its centered position in the beam.  These shifts did not produce any changes in the neutron yields, indicating that the neutron beam is confined to the central region of the samples.  Previous experiments~\cite{frisch46,bailey46}, in which the beam profile was larger than the detector, required an additional correction due to neutrons produced at angles outside the detector acceptance which scattered into the detector.  In our geometry, this correction was not required.

The bombarding proton energies ranged from 2.00 to 2.55 MeV in 50 keV increments, yielding neutrons from 150 to 800 keV.  This spacing produced minimal overlap of neutron energies, especially above a proton energy of 2.25 MeV.  This resulted in an increase in the statistical uncertainty near neutron energies of 650, 700, and 750 keV.  For proton energies above 2.25 MeV (neutron energies above 450 keV), a total of four hours of data were taken at each proton energy.  For proton energies below 2.25 MeV (neutron energies below 450 keV) a total of eight hours of data were taken at each proton energy.  These additional statistics were collected to compensate for the decreasing $^7$Li$(p,n)^7$Be cross section and the increasing $np$ total cross section.

Four polyethylene and three carbon samples were used, in addition to the empty position used to determine the yield with no sample in the beam.  A caliper was used to measure each target dimension with an uncertainty of 0.25 mm.  The mass was determined with a scale to an accuracy of 10 mg. The polyethylene samples were rectangular prisms, with nominal length and width of 7.6 and 5.1 cm, with a thickness varying from 0.5 to 3.0 cm.  The carbon targets were cylindrical, with a diameter of 4.8 cm and a thickness varying from 3.1 to 6.1 cm.  The nominal thickness and areal density of each sample are given in table~\ref{tab:jan11targets}. 
\begin{table}
\centering
\begin{tabular}{ c | c | c }
Material & Nominal Thickness (cm) &  $\tau$ (g/cm$^2$)  \\
\hline
CH$_2$ & 0.5 & $0.479\pm0.003$\\
CH$_2$ & 1.0 & $0.959\pm0.006$\\
CH$_2$ & 2.0 & $1.93\pm0.01$\\
CH$_2$ & 3.0 & $2.91\pm0.02$\\
C & 3.1 & $4.89\pm0.02$\\
C & 4.6 & $7.24\pm0.03$\\
C & 6.1 & $9.42\pm0.04$
\end{tabular}
\caption{CH$_2$ and C sample thicknesses.\label{tab:jan11targets}}
\end{table}
Approximately 30\% of the transverse area of each sample was exposed to the neutron beam.

The samples were mounted on a remotely-controlled wheel which could position up to twelve samples in the neutron beam.  During each two hour run, the wheel would automatically switch which target was in the beam at pre-programmed intervals, resulting in a series of irradiations between one and seven minutes long.  Feedback signals from the wheel indicated when samples were being moved, and which sample was in the beam at a given moment.  By using these short intervals, the yields from each irradiation of the samples could be compared under similar beam conditions, eliminating potential variations due to beam current fluctuations or the condition of the LiF target.  At each of the twelve positions, the samples were mounted onto a smaller wheel.  When in the beam position, a second motor rotated this smaller wheel so as to average over possible non-uniformities in thickness.

\begin{figure}
\centering
\includegraphics[width=7cm]{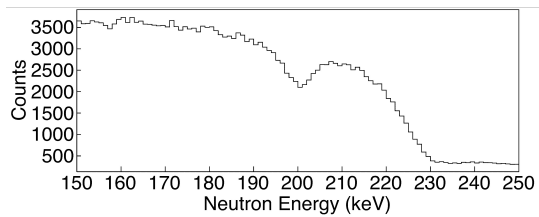}
\caption{Neutron energy spectrum with a sulfur sample for $E_p=2.00$ MeV.  This corresponds to a maximum neutron energy of 230 keV, and there is a visible decrease in counts associated with the sulfur resonance at $E_n=203$ keV.}
\label{fig:3}
\end{figure}

Additionally, a sulfur sample was used to calibrate the neutron energy.  With its multiple resonances in our energy range, we can use the observed decreases in yield and the maximum neutron energy for each proton beam energy to calibrate the absolute time of flight.  The neutron energy spectrum with a sulfur sample for $E_p=2.00$~MeV is shown in figure~\ref{fig:3}.

The data were recorded event by event, using NIM and CAMAC electronics.  The dead time ranged from 5-40\%, depending on the event rate.

\section{Analysis}

\begin{figure}
\centering
\includegraphics[width=7cm]{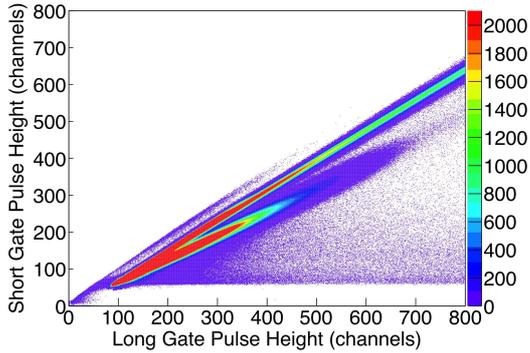}
\caption{(Color online) Neutron detector ADC outputs: short-gated pulse height versus long-gated pulse height.  The upper broad band is $\gamma$-rays, the lower band is neutrons.}
\label{fig:4}
\end{figure}

\begin{figure}
\centering
\includegraphics[width=7cm]{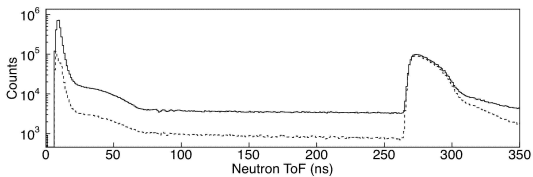}\\
\caption{Neutron time of flight for $E_n=450$ to 500 keV with no conditions (solid line) and the pulse shape discrimination condition (dashed line).  The $\gamma$-flash is visible at 10 ns, with the neutron peak visible at 270-300 ns.  With the pulse shape discrimination condition, the background is reduced by 75\% outside the neutron peak.}
\label{fig:5}
\end{figure}

We used pulse shape discrimination in the neutron detector to separate neutrons and $\gamma$-rays, thus significantly reducing the background.  The pulses from the liquid scintillator were split and were recorded in separate ADCs using a long gate (500 ns) and a short gate (100 ns).  The short-gated pulse height versus the long-gated pulse height is shown in figure~\ref{fig:4}.
The upper and lower broad bands correspond to $\gamma$-rays and neutrons, respectively.  They merge at small pulse heights, where the difference in the long and short portions of the pulse is too small to separate, but using a conservative cut allows the elimination of all higher energy $\gamma$-rays.  Figure~\ref{fig:5} shows the neutron time of flight spectrum for $E_n=450$ to 500 keV with no conditions (solid) and the pulse shape discrimination condition (dashed).  The background outside the neutron peak (270-300 ns) is reduced by 75\% while the neutron yield is unaffected.
Neutron yields were determined in 10 keV bins from 150 to 800 keV, and the integrated live current was recorded for each sample irradiation in order to normalize the neutron yields.

To extract a cross section from the yields from each sample, we normalized the neutron yields by the proton beam current,
\begin{eqnarray}
\sigma _C &=&\frac{1}{\tau _c} \text{ln}\frac{(Y/Q)^{\text{out}}}{(Y/Q)^{\text{C}}},\label{eq:carbonsigma}\\
\sigma _p &=&\frac{1}{\tau _p} \text{ln}\frac{(Y/Q)^{\text{out}}}{(Y/Q)^{\text{CH$_2$}}}-\frac{\sigma_c}{2},\label{eq:protonsigma}
\end{eqnarray}
where $\tau_p$ and $\tau_c$ are the sample thicknesses from table~\ref{tab:jan11targets}, $Y$ is the neutron yield, and $Q$ is the integrated live-time current.  The carbon contribution of the polyethylene targets is subtracted, and the uncertainty in the $n$C cross section is included in the total uncertainty in the $np$ cross section.  Due to the long lengths of the CH$_2$ chains, the variation of the hydrogen to carbon ratio from 2 is insignificant compared to the other systematic uncertainties.  These results do not require knowledge of the neutron detector efficiency nor the absolute beam flux.  The proton beam current was integrated for each irradiation, and the dead time of the data acquisition system was precisely measured as a function of the event rate in order to determine the total live-time current.

The backgrounds were divided into two categories: the room background, which is independent of the time of flight, and the sample background, which is due to neutrons which scatter from the sample but are still detected.  The room background can be measured from the data by fitting to the constant background outside the neutron time-of-flight peak.  The sample background was determined by extrapolating to zero thickness.  For each target, the cross sections measured at each energy were combined using a weighted average, in order to produce a single average cross section for the entire energy range.  These average cross sections were then plotted versus target thickness and fitted with a linear function.  The intercept of this linear function gave the average cross section at zero target thickness.  By taking the ratio of the average cross section at zero thickness and the average cross section for each sample, we determine the correction for the background due to target thickness.   These corrections ranged from 0.3\% for the thinnest sample to 1.3\% for the thickest sample.  The corrected cross sections for each energy bin were then combined using a weighted average.

\section{Results}

\begin{figure*}
\centering
\includegraphics[width=\textwidth]{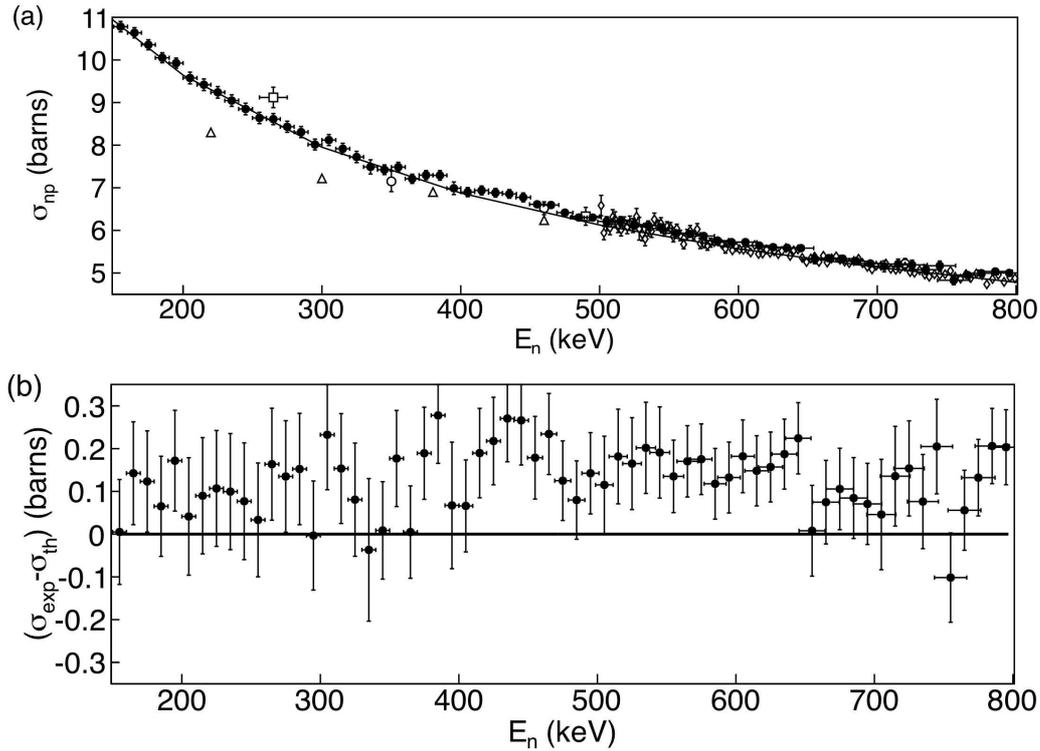}
\caption{(a) Results for (filled circles) the $np$ total cross section measurements for neutron energies from 150 to 800 keV, plotted with ENDf/B-VII.1 tabulation (solid line) from Hale et. al.~\cite{hale04} and previously measured $np$ cross sections from 150 to 500 keV, measured by (squares) Frisch~\cite{frisch46}, (open circles) Bailey et. al.~\cite{bailey46}, (triangles) Bretscher et. al.~\cite{bretscher50}, and (diamonds) Clement et. al.~\cite{Clement197251}. (b) The difference between our experimental cross section and the ENDf/B-VII.1 tabulation.  Error bars include the statistical and systematic errors added in quadrature, along with the contribution due to subtracting the experimental $n$C cross section.}
\label{fig:6}
\end{figure*}

\begin{figure*}
\centering
\includegraphics[width=\textwidth]{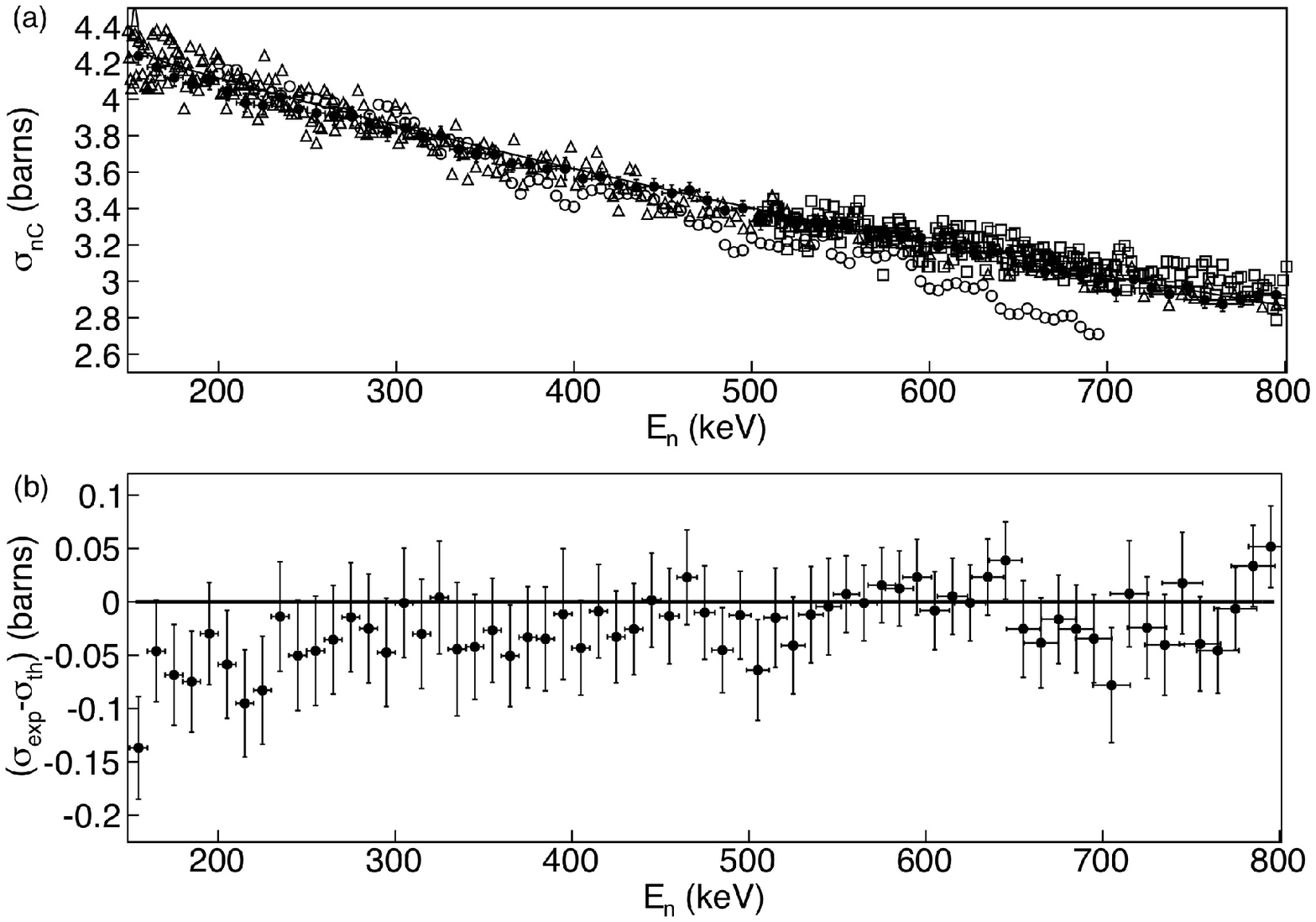}
\caption{(filled circles) Results for the $n$C total cross section measurements for neutron energies from 150 to 800 keV, plotted with the ENDf/HE-VI tabulation~\cite{pearlstein93} and existing data, measured by (squares) Huddleston et. al.~\cite{huddleston60}, (open circles) Wilenzick et. al.~\cite{wilenzick61}, and (triangles) Uttley et. al.~\cite{uttley64} Error bars are suppressed in the previous $n$C measurements for clarity in the plot.  (b) The difference between our experimental cross section and the ENDf/HE-VI tabulation.  Error bars include the statistical and systematic errors added in quadrature.}
\label{fig:7}
\end{figure*}

Our results are shown in figure~\ref{fig:6}a for the $np$ total cross section and in figure~\ref{fig:7}a for the $n$C total cross section.  Existing $np$ data~\cite{frisch46,bailey46,bretscher50,Clement197251} and the ENDf/B-VII.1~\cite{hale04} tabulation are also shown in figure~\ref{fig:6}a, and existing $n$C data~\cite{,huddleston60,wilenzick61,uttley64} and the ENDf/HE-VI~\cite{pearlstein93} tabulation are also shown in figure~\ref{fig:7}a.   Figures~\ref{fig:6}b and~\ref{fig:7}b show the difference between our experimental cross sections and the tabulations.
Error bars are suppressed in the previous $n$C measurements for clarity in the plot; the uncertainty in these measurements is between 2.7\% and 5\%.

The total uncertainty in each of our measurements is  in the range 1.1-2.0\%.  The systematic uncertainties included the measured masses and dimensions of the samples, listed in table~\ref{tab:jan11targets}, contributing 0.35\%; the characterization of the beam and system deadtime, contributing 0.5\%; and the background determination, contributing 0.5\%.  The statistical uncertainty was of the order of 0.4\%.  The uncertainties due to the target dimensions applied uniformly to all data points, while the statistical, beam, system dead-time, and background uncertainties were determined for each sample irradiation.

\section{Discussion}

In order to connect our measurements to $NN$ potential models, we parameterize the $s$-wave neutron-proton elastic scattering cross section,
\begin{equation}
\sigma = \frac{3}{4}\sigma_t + \frac{1}{4}\sigma_s,
\end{equation}
in terms of the triplet and singlet scattering lengths $a_{t,s}$ and energy-dependent effective ranges $\rho_{t,s}(0,T)$:~\cite{sachs,hackenburg06}
\begin{equation}\label{eq:swave}
\sigma_d=\frac{4\pi}{(a_d^{-1}-\frac{1}{2}\rho_d(0,T)p^2)^2+p^2)},
\end{equation}
where the subscript $d$ represents either $t$ or $s$, $\rho_d(0,T)$ is the energy-dependent effective range,  and $T$ and $p$ are the center of mass kinetic energy and momentum, respectively.  Hackenburg~\cite{hackenburg06} has recently revisited the problem of determining the zero-energy cross section and effective-range theory (ERT) parameters from data, including a consideration of the correlation between the singlet and triplet effective range.  Measurements of the zero-energy $np$ cross section, $\sigma_0$~\cite{melkonian49,houk71}, and the parahydrogen coherent scattering length, $a_c$~\cite{koester71,koester75}, given by
\begin{equation}
\begin{array}{rcl}
\sigma_0 & = & \pi(3a_t^2+a_s^2),\\
a_c&=&\frac{3}{2}a_t +\frac{1}{2}a_s,
\end{array}
\end{equation}
were included in the fit due to correlation between $a_s$ and $a_t$.  The ERT parameters resulting from the fit were $\rho_t (0,0) = 1.718 \pm 0.025$ fm and $\rho_s (0,0) = 2.696 \pm 0.059$ fm.  Also, the zero-energy shape dependence parameter $\Delta r_t = -0.025 \pm0.025$ fm.  Hackenburg's analysis also concluded that additional total cross section measurements for $np$ scattering between 20 and 600 keV were required in order to use the parameters of effective range theory to differentiate among $NN$ potential models.

It was hoped that the present results in the 150-600 keV energy range would reduce the uncertainties in the parameters determined from the fit to the ERT expression (Eq.~\ref{eq:swave}) for the cross section.  Following Hackenburg's method, we included our data in the fit.  The resulting parameters were consistent with those determined by Hackenburg.  The uncertainties were not significantly reduced, as our measurements were not precise enough to further constrain the parameters.  According to Hackenberg, a 0.004\% precision measurement at 130 keV is required to reduce the uncertainty in $\rho_t(0,0)$ and $\Delta r_t$ to 0.001 fm~\cite{hackenburg06}.  By instead measuring the energy dependence of the cross section, we estimate that a precision of 0.5\% across the 150 to 800 keV range currently measured is required to decrease these uncertainties, and 0.1\% measurements across this range will decrease the uncertainty to 0.01 fm.

\section{Conclusion}

Our measurement of the $np$ total scattering cross section has filled in a large gap in the total cross section measurement below neutron energies of 500 keV.  By measuring ratios of transmitted events with and without the samples in the beam, we were able to determine the cross section independently of the neutron detection efficiency.  Both the $np$ and $n$C total cross sections are consistent with previous measurements, and the $n$C results show significantly decreased scatter and uncertainty.

There is a slight systematic discrepancy between our measurements and the ENDf tabulations, with a $-18$ mb shift downward in the $n$C cross section, and a $+136$ mb shift upward in the $np$ cross section.  This discrepancy would not have been visible in the previous $n$C data.  The previous $np$ data in this range also show a slight increase relative to the theoretical curve.

However, when the present data are used to try to improve the fit of effective range theory, we find that our measurements are not precise enough to increase the precision on the resulting parameters.  The effective range parameters $\rho_d$ are most sensitive to measurements in the 20-600 keV region, but more precise data at higher energies have already constrained the parameters more narrowly than the precision of current measurements in this region.  The discrepancy noted previously is comparable to the current uncertainty, and so does not significantly impact the determination of the ERT parameters. To be comparable to the current uncertainty in $\rho_t(0,0)$ and $\Delta r_t$, 0.5\%  precision would be required across the energy range measured here, 150 and 800 keV.  This precision is attainable with increased statistics and better characterization of the beam, both of which are possible with the current configuration.

\section*{Acknowledgments}

The authors would like to thank J. French for her work in the setup and running of the experiment.  This work was supported by the United States Department of Energy, the National Science Foundation, and the SSAA.

\bibliography{../../references}{}

\end{document}